# Title

- Coherent chiral computation without optical parametric oscillation.

# Authors

Yongpan Gao,[1]* Pengfei Lu,[1] Chuan Wang[2]†

# Affiliations

- [1] State Key Laboratory of Information Photonics and Optical Communications and School of Electronic Engineering, Beijing University of Posts and Telecommunications, Beijing,100876,China
  [2]School of Artificial Intelligence, Beijing Normal University Beijing,100875,China
- *Correspondence to: gaoyongpan@bupt.edu.cn
  †Correspondence to: wangchuan@bnu.edu.cn

# Abstract

The challenge posted by modern science is to find a way to compute the NP-hard problem. Here we present a coherent computation model based on the whispering-gallery mode micro-resonators. We introduce the optically connected micro-resonators array to simulate the dynamical evolution of the Ising model with full connection while the mode chirality of the micro-resonators works as the spins of each node in the Ising model. The relaxation time, the steady state distribution, and the mode connection of the Ising spin chain are simulated with the proposed optical system. Moreover, it is found that the current scheme could be used for searching the full solutions of the corresponding Ising problem. Numerical results show that the time required is constant time. Our approach achieves high accuracy and fast computation in the realization of unconventional optimization hardware.

# Teaser

Ising computer based on the rotation directionality and nonlinearity in an optical whispering gallery mode microcavity.

# MAIN TEXT

## Introduction

The main challenges encountered in computer science is to find the effective solution of nondeterministic polynomial (NP) problems. Due to the rapid progress of quantum computation (1-3), various approaches have been employed to solve this problem, such as the quantum annealing, quantum simulate machines, and coherent Ising machines. It can be proved that some NP problems can be physically solve in linear time complexity. For example, recent studies suggest that the quantum annealing (QA) which is to find the global minimum of a given objective function could be used for solving the combinatorial optimization problems with several local minima (such as finding the ground state of a spin glass (4) or the traveling salesman problem ).

Specifically, the spins in such Ising systems enable approaches to solve the sort of NP problem, e.g. the combinatorial optimization problems. The Ising model is a standard and important model in statistical physics. It consists of discrete variables of spins that represent the magnetic dipole moments of electrons' spins in one of two states which is originally used to describe the ferromagnetism of spin chains in statistical mechanics (5-7). Recently, it has been investigated that the computation of its ground state is similar to a search problem which is non-deterministic polynomial-time hard . As the Ising model without an external field is equivalent to the of graph maximum cut(MAX-CUT) problem in combinatorial optimization. Equivalently, other related combinatorial optimization problems, such as the cost functions, the minimum spanning tree problem, and the traveling salesman problems, can also be solved by this model(8-12).

As an optical Ising simulator, the coherent ising machine(CIM) based on the degenerate optical parameter oscillation(DOPO) has recently received extensive attentions(11-19). CIM is originally proposed to simulate the Ising model with the optical parametric oscillation process(11), which could solve the max-cut problem in largescale(8-12). Following an intuitive approach, DOPO could mimic the working principle of spin up and spin down states by choosing different optical in-phase and out-phase amplitudes of the femtosecond laser pulses. Technologically, the energy efficiency of the DOPO system is lower than 17%(20,21), and the scale of photonic qubits is limited by the repetition rate of the femtosecond lasers. Therefore, although the energy efficiency of the current scheme is low, it is computational complexity could be approaching $O(n)$. Later, this feature has achieved several exciting results in solving of the MAX-CUT problem (15,16). Meanwhile, various prototypes optical system are been employed to simulated the Ising mode, such as the injection-locked lasers(14,22-25), and the spatial light modulator (SLM)(26-29).

The CIM is mainly constructed in the optical system. In optical systems, especially integrated optical systems, microcavity is a important structure. Optical microcavity have the properties of light localization and enhancement of light-matter interaction(30-32). Many nanoscale quantum optical effects can be exploited for a variety of applications in this field, such as photon blockade(33-35), light medicated matters interaction(36-39), frequency comb(40-45), high sensitivity sensing (46-49), singular phenomena in non-Hermitian system (47,50-52), chaos assisted transmission (53-56), and so on. The use of optical microcavities to realize the optical Ising structure is of great significance to the development of microcavity optics and coherent Ising machines(CIM) calculations. For microcavity optics, researchers nowadays discuss its properties mainly on the control of physical phenomena. For CIM, the previous solutions can often only achieve one of the characteristics of low threshold, full connectivity, and high energy efficiency. The comprehensive development of the characteristics of the microcavity and the construction of CIM can not only endow the optical microcavity with new functions, but also the key to achieving a low threshold, full connection, and high energy efficiency CIM.

Here we introduce a representation of the coherent computation model in terms of optical Ising chain specified by a set of full connected WGM micro-resonators.

## Results
**Spins configuration in optical whispering-gallery modes microresonators.**

Here we present a programmable Ising model with whispering-gallery mode (WGM) microcavities. Due to the scattering interaction, the WGM cavity could support two degenerate optical modes: the clockwise(CW) mode and the counter-clockwise(CCW) mode. It is found that such a classical bifurcation could work as an Ising machine in which the two modes could simulate the spins in the Ising model. We further present a fully connectable Ising model without optical wavelength conversion. The principles of the model is shown in Fig.1. Specifically, the Fig.1(a) shows the CW and CCW modes balanced pumping scheme in a single WGM microcavity. Consequently, we choose each mode amplitude of the bifurcates to a positive or negative value. The pumping strength of these two modes fulfills the relationship $a_{cw,in} = a_{ccw,in}$ and they are linear scattered and coupled in the Kerr form(57-59).

The full scheme of the Ising machine is shown in FIG.1(b), i and j are denote the node numbers of the Ising model. The output of the cavity field is collected by the Programmable Coupling Component(PCC). The PCC generate its output with the collected results, preset the program and give an output $b_{CW(CCW),i}$. Subsequently, the output $b_{CW(CCW),i}$ together with the continuous wave laser input is injected into the corresponding cavity. The PCC can be achieved with programmable photonic integrated circuits(60).

The theoretical model for a single cavity could be established by presenting the mode dynamical equation as(58).

$$\frac{da_{CW}}{dt} = -\frac{\kappa_1}{2}\frac{a_{CW}+a_{CCW}}{2} - \frac{\kappa_2}{2}\frac{a_{CW}-a_{CCW}}{2} + i\Delta a_{CW} + iga_{CCW} + igM\left(|a_{CW}|^2 + 2|a_{CCW}|^2\right)a_{CW} + \sqrt{\kappa_{in}}a_{CW,in}$$
(1a)

$$\frac{da_{CCW}}{dt} = -\frac{\kappa_1}{2}\frac{a_{CCW}+a_{CW}}{2} - \frac{\kappa_2}{2}\frac{a_{CCW}-a_{CW}}{2} + i\Delta a_{CCW} + iga_{CW} + igM\left(|a_{CCW}|^2 + 2|a_{CW}|^2\right)a_{CCW} + \sqrt{\kappa_{in}}a_{CCW,in}$$
(1b)

Here $a_\mu$ denotes the amplitudes of mode μ = CW; CCW, and the $\kappa_i$ is the total damping rate of the i-th node. Δ is the drive-cavity frequency detuning. g represents the scattering coupling strength of the CW and CCW modes. M symbols the Kerr nonlinear factor, and. $\kappa_{in}$ is the coupling strength between the cavity mode and the optical fiber. For a single optical cavity, the relation between the spin of Ising model and the WGM modes can be created as：

$$\sigma_z = \begin{Bmatrix} a_{CW} & 0 \\ 0 & a_{CCW} \end{Bmatrix} \quad (2)$$

When there is non-negligible nonlinearity in the system, and the mode intensity approaches the threshold, one of the two modes will approach the zero value randomly due to the nonlinearity. If we set the logic value 0(1) related to the weaker (stronger) amplitude, the $\sigma_z$ matrix will have the same form as spin up(CW) and spin down (CCW) states.

To qualitatively explain the operation principle of the scheme, we numerically simulated the system by solving the ordinary differential equations(ODE) Eq.1 with randomly chosen initial values, and present the results in Fig.2. All the parameters in our simulation are experimentally achievable, and the parameters and variables of the system are associated with the dimensionless quantities. In detail, the parameter M is $3.64 \times 10^{11}$ $J^{-1}$ and the damping rate is chosen as $\kappa_1=\kappa_2=11.6$ MHz, the coupling strength is set as $g=2.93$ MHz. We set the parameter $\hat{a}$ is $10^6$ $\hat{a}$ in the simulation. Under the replacement, all parameters in the equation is about 1 unit (in μs). In Fig.2(a), we perform numerical calculations of Eq.3 under different detuning($\Delta$), and choose its steady state as the transmission axis. We found the simulated transmission spectrum is well consistent with the previous experiment results (58). In Figs. (b) and (c), we verify that the random features of the CW and CCW modes of the microcavity. Here the red line denotes the CW amplitude, while the blue line represents the CCW amplitude. Here we mark the stronger amplitude with logic-1, while the lower amplitude with logic-0, and Fig.2(b) shows the CCW logic-0 state and Fig.2(c) shows the CW logic-0 state. This symbols that the numerical simulation can reproduce the results of the random state in the experiment. The numerical simulation is performed by 100 times and the state distribution is shown in Fig. 2(d). We can find in the results that the CCW logic-0 state is 48 times, and the CW logic-0 state is 52 times, which is also consistent with previous experiment results(58.61-63)

Here, we show an example of the Ising model dynamics using the single WGM microcavity as above. Since it is consistent with the experiment using single cavity, we generalize it to explore the multi-cavity situation. When the cavities coupled with each other, the dynamical equations could be expressed as

$$\frac{da_{CW,i}}{dt} = -\frac{\kappa_1}{2}\frac{a_{CW,i}+a_{CCW,i}}{2} - \frac{\kappa_2}{2}\frac{a_{CW,i}-a_{CCW,i}}{2} + i\Delta a_{CW,i} + iga_{CCW,i} + igM\left(\left|a_{CW,i}\right|^2 + 2\left|a_{CCW,i}\right|^2\right)a_{CW,i}$$
$$+\sqrt{\kappa_{in}}a_{CW,i,in} + \sum_{j\neq i}J_{i,j}a_{CW,j}$$
(3a)

$$\frac{da_{CCW,i}}{dt} = -\frac{\kappa_1}{2}\frac{a_{CCW,i}+a_{CW,i}}{2} - \frac{\kappa_2}{2}\frac{a_{CCW,i}-a_{CW,i}}{2} + i\Delta a_{CCW,i} + iga_{CW,i} + igM\left(\left|a_{CCW,i}\right|^2 + 2\left|a_{CW,i}\right|^2\right)a_{CCW,i}$$
$$+\sqrt{\kappa_{in}}a_{CCW,i,in} + \sum_{j\neq i}J_{i,j}a_{CCW,j}$$
(3b)

This equation described the model which can be connected with single case in Fig.1, and the connection coefficient between different cavities is denoted as $J_{i,j}$. Here the value $J_{i,j}$ corresponds to the Ising coupling strength. As discussed in the above that the single nonlinear cavity have features of the single spin, the coupled cavities could also exhibit the features of the Ising model. In fact, the coupling between different cavities is similarly with the coherent Ising machines (11).

We choose all the cavities in the scheme have the same parameters as the single cavity condition. In addition, the coupling strength between the modes in different cavities is set as $J_{i,j}=0.1$. In Fig.3, we show the dynamical evolution and the state probability statistics. Under normal circumstances, two cavities should have four possible states with the logic vector $a_{cw,1}$ $a_{cw,2}$ which are (0,0), (0,1), (1,0), and (1,1). However, the simulation shows that only two possible states exists, say the (1,0) and (0,1) states.

Due to the coupling between the input waveguide and the cavity, the input of the CW and CCW modes for the same cavity is no longer balanced, and the states between different cavities could affect each other. Here Fig.3 shows two possible modes in the coupled cavity system. In order to fully describe the nature of the system, we set the two cavities under weak coupling g = 0:05, see Fig.3(c). The probability that we get 01 and 10 states are both 50% probability There is also a little probability one can get (1,1) and (0,0) state, it is because the two cavity may research it steady state at the same time.

It is noticed that the higher amplitude modes should be focused. By recording the CW mode or CCW mode, then we can get a vector spanned by the matrix $(CW, CCW)^T$. If the CW value is assigned as 1 and CCW value is 0, we can get write the final state of the system as (1,0). The above two assignment methods are equivalent, as only one of the CW and CCW is stronger than the other relative weak one for the same cavity.

Meanwhile, the operation time, i.e. the relaxation time, is also a key parameter. In the following part, we show the performance of the relaxation time under different driven strength in Fig.3(d). The results presented here are counted for the 100 times simulations at each point in the figure. By increasing the pumping power in the unit from 1.6 to 1.66, we can find that the relaxation time decreases as the increment of the pumping power upon the threshold value. For example, when the pumping power is 1.60, the relaxation time is 250 , and when the pumping power is increased to 1.66, the pumping power decreases to 50. Even more, in terms of the deviation of the average value from other values, the relaxation time of the system is more unstable under weak pumping condition.

The Ising model in our scheme is full connected. However, there are two ways of coupling when the connection between different cavities is constructed: all the CW modes are only coupled with the CW mode (CW-CW coupling mode); and all the CW modes are only coupled with the CCW mode (CW-CCW coupling mode). For the CW-CW mode coupling, the dynamical behavior is shown in Eq.3; while for the CW-CCW coupling, we should first exchange the index CW and CCW in the last term of Eq.1. These two coupling conditions are the same for two-cavity condition but different when cavity numbers exceed three for the symmetry broken of the CW and CCW modes under the coupling. For simplicity, we choose the three-cavity condition as an example. By selecting the same parameters as the two cavities coupling condition, we show the statistical simulation results of circular coupling three optical cavities in Fig.4 under 300 sample.

For the sake of simplicity, we use a string of 0 and 1 to represent the state of the cavities sequence. The logic-1(0) stands for the CW(CCW) mode is stronger than the CCW(CW). For example, the sequence of 101 represents that the CW mode is stronger in cavity-1, CCW mode is stronger in cavity-2, and CW mode is stronger in cavity-3. In Fig.4, we can find the system can only be one of the 101, 110, 011, 010, 001, and 100 state. Even more, we found all these states have almost the same probability, so we can get all possible Ising ground states with equal probabilities.

The equivalence between the system and the Ising model provides an effective way to solve the combinatorial optimization problems, such as the MAX-CUT problem (The full discussion can be seen in Supplementary materials). In Fig.5, we shows all possible simulation results of a four-cavity coupling system, in which the first three cavities are mutually coupled, and the forth cavity is only coupled with the third cavity. And all these cavities coupling in the anti-$\sigma_z$ form. We only show the amplitude of the CW mode in this figure and the following. Here, the blue line, the red line, the green line, and the pink line indicate the time-dependent evolution of the CW mode of the first to fourth cavities, respectively, during the simulation of the MAX-CUT problem. Then, we can solve the solution of the problem and get the results are 1011, 0100, 1010, 0101, 0110, 1001 from Fig.5(a)-f. And these are all the related solutions to the corresponding MAX-CUT problem. So this model provides a MAX-CUT solver that can give full solutions.

To suppress analog errors in practical applications, we also verify the effectiveness of the system under a large number of nodes. To show the validity of the scheme for large scale system, the dynamical evaluation of the system with the cavities coupled in a chain form is numerically simulated, i.e. the i-th cavity only coupled with the i-1-th and i + 1-th cavity. We simulate the Ising chain with 16, 32, 64, 256, and 1024 cavities(nodes) in Fig.6 (a), (b), (c), (d), and (e), respectively. All the parameters are chosen the same as in the previous figure, except that we reduce the dissipation of the cavity to 0.01 in Fig. (b) (c) (d) and (e). For simplicity, we use the red line to symbol the even number cavities' CW mode amplitude, and blue line to symbol the odd number cavities' CW mode in (a), (b), (c), (d). Here the number corresponds to the position of the cavity in the Ising chain. For a chain connected MAX-Problem, its solution is to separate the nodes according to the numbered parity, represented by a vector (1; 0; 1; 0…….).

In Fig.6(a) , the parameters are the same as in Fig.5, while in (b), (c), and (d), the coupling strength between different cavities is set to 1 unit. The simulation results in these four figures prove that we can successfully get the right result under various conditions. As the size of 256 bits may not be large enough, we show the 1024 cavities steady state of all CW and CCW modes in Fig.5(e), here different colors represent different amplitude of the modes . There are obvious color bands in the picture which show that the odd-numbered cavities and the even numbered cavities are well separated and the corresponding 1024 nodes MAX-CUT problem could be efficiently solved. In short, the above simulation proves that our model is also valid for 1024 bits, which is comparable with the newest result of the DOPO scheme in 2016(15)

Next, we studied the scaling of the computation time of the system. We can get the time complexity through the relation time in Fig.6. The damping rate is 0.1 for (a), while it is 0.01 for (b), (c), and (d). Although the number of nodes is different, the stable time is only relate to the damping of the system. So the computation time and the complexity of the WGM Ising scheme is O(1). This can be attributed to the synchronous drive allows all nodes to enter the relaxation process in parallel, so the steady state time of the system is independent to the number of nodes, but only dependent on the damping rate of each cavity, i.e. O(1) time complexity.

At last, we compared the accuracy of the solutions delivered by the system. As the number of nodes increases, the correspondence between the nonlinear coupled system to the Ising model shows instability, and the errors would appear. To deeply understand the features of the error situation, we show the failed examples for the 128 bit and 256 bit condition In Fig.7, respectively. The figure provides that there is an overlap between the odd modes and even modes in the steady region which means the failure of the calculation.

In fact, the errors are inevitable in the computation. To reduce the influence of the error, error correction process is required. Fortunately, there is a natural error correction scheme in our system. When we continue to explore the properties of the Fig.7, we can find that when the errors are produced, there will be a sub-weak amplitude modes near the logic-0 mode. Its formation has the same mechanism as the domain wall in the magnet, we call it a domain wall error. Compared with the above successful calculation in Fig.6, the system mode under successful calculation can only have two strengths. This domain wall could be employed to mark incorrect calculations. When the system has a sub-weak mode(domain wall) close to logic 0 in the steady state of the system, the calculation result could be marked as an error and we ran the calculation again

**Conclusion**

In summary, we present an optical chirality simulation of Ising model and programmable coherent computation protocol with WGM optical microcavities. The clockwise and counterclockwise modes of the WGM are employed to construct the Ising spin up and spin down state. We achieved the full solution search MAX-CUT problem in the O(1) times. In our simulation process, the maximum cut problem with the scale 1024 bits is achieved. For possible calculation errors, we proposed a method to realize the error judgment by whether the system has a sub-weak mode (domain wall) close to the weakest mode.

Even more, the energy efficiency of this Ising model is 100%. At the output, the system can be implemented directly through the photoelectric converter, the complexity of the detection port is greatly reduced. Our system proposes a novel and stable optical simulation method of the Ising model and create a O(1) time complexity 100% energy efficiency MAX-CUT solver. We believe our method can be further used as an excellent on-chip platform for non-classical optimization problem.

**Materials and Methods**

In order to avoid errors caused by approximation, all conclusions are based on the results of numerically solving differential equations. We solved the equations in this article with implicit Runge-Kutta method. In order to reduce the amount of calculation, we use standard unit values in the numerical calculation process, but set 1 microsecond as a time unit according to the evolution of the system. The maximum time step we used in the numerical calculation process is 0.001, which is 1ns. In order to present the statistical results of Figure 2(d), Figure 3(c-d), and Figure 4, we randomly assume that the initial value of the system is 0.01 intensity units, which is far less than the stable intensity of the system. What we need to add is that when the step size of our numerical simulation is different, the system will also be in a different steady state.

**Acknowledgments**

**Funding:** We would like to thank the support from National Natural Science Foundation of China under Grant Nos. 62131002 and 62101057.
**Competing interests:** The authors declare that they have no other competing interests.
**Data and materials availability:** All data needed to evaluate the conclusions in the paper are present in the paper and/or the Supplementary Materials. Additional data and codes related to this paper may be requested to the authors.

**Figures and Tables**

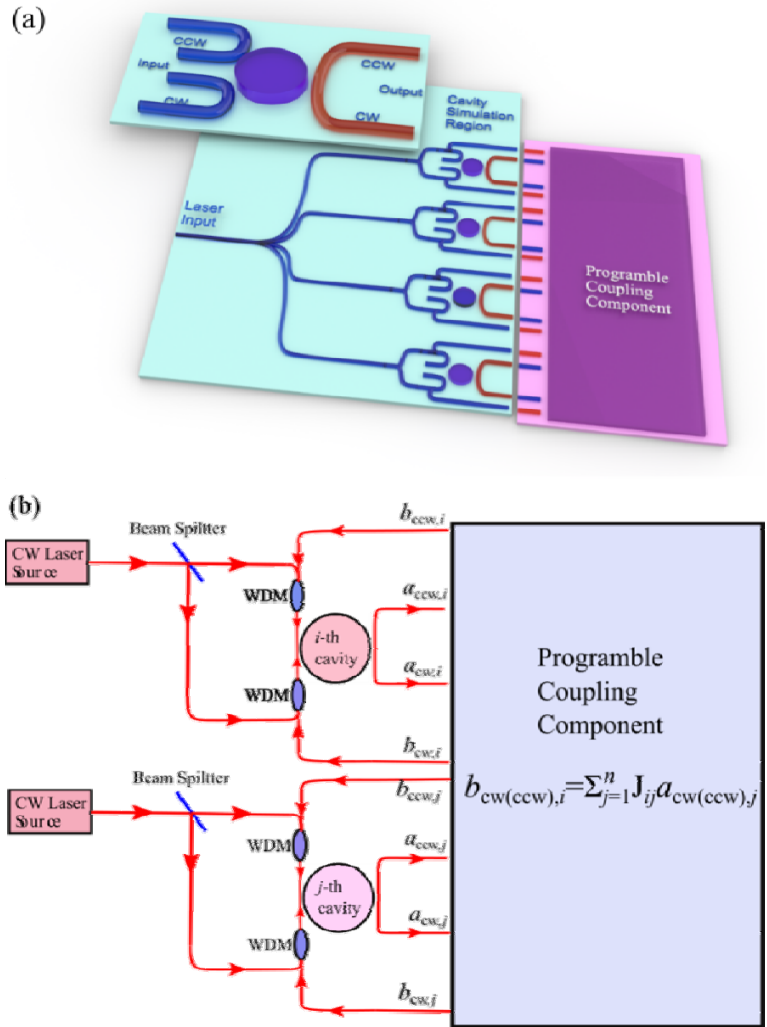

**Fig. 1. The scheme of the chirality Ising dynamics.** (a)The single cavity mode is used to simulate the single Ising spin. (b) The scheme of the programmable coupled cavities. We set the *i*-th and *j*-th cavities as an example. The CW and CCW modes in each cavities are driven by a continuous wave laser with the same power. Then, the output field of the cavity is detected with the Programmable Coupling Component (PCC) and output a feedback $b_{CW,j}$. The amplitude $b_{CW(CCW),j}$ together with the laser drive are then coherently input into the cavity again.

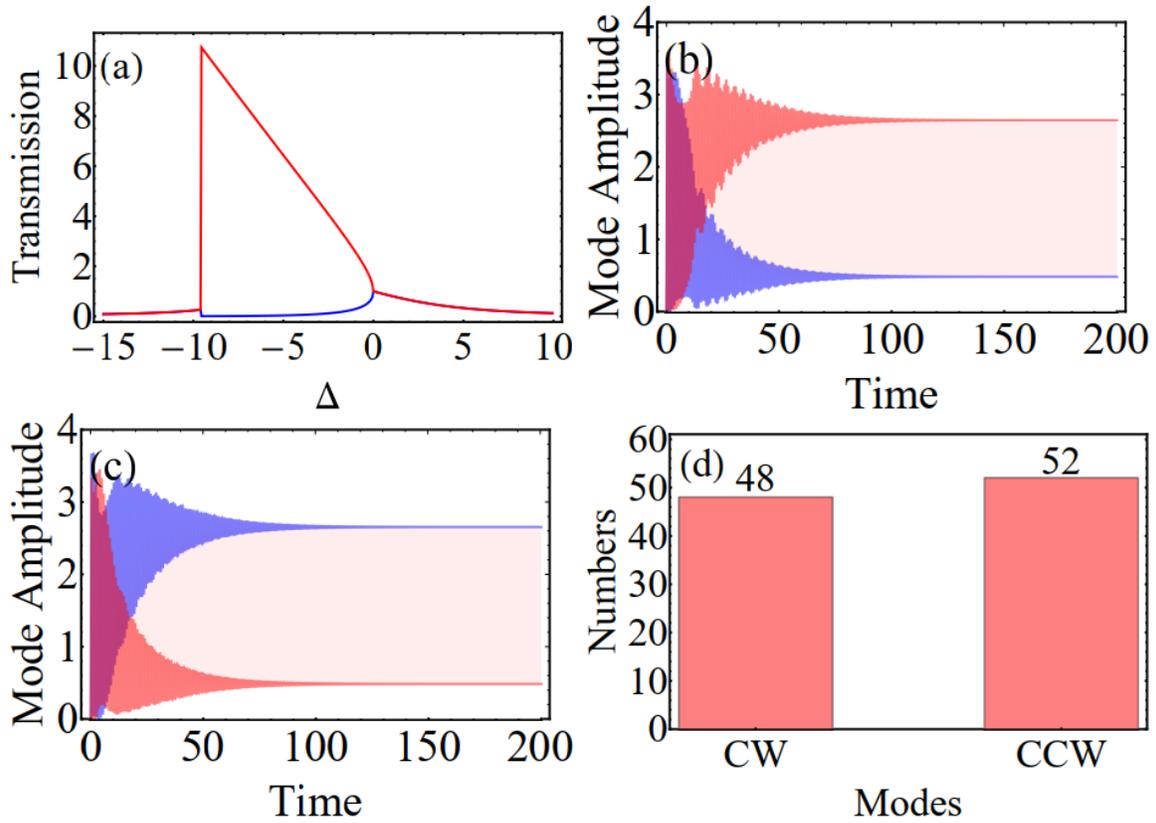

**Fig. 2. The numerical simulation of the steady state amplitude in the silica WGM cavity.** The system is simulated with the pumping detuning chosen as Δ=-1, the dissipation is 0.1, the pumping power is 5. The scattering strength of the field is set as g=1, the nonlinear coefficient is M=1. (a) The transmission spectrum results versus the detuning. (b) and (c) are the two possible results under the balanced drive. (d) The statistical distribution of the two results (CW or CCW is stronger) in (b) and (c).

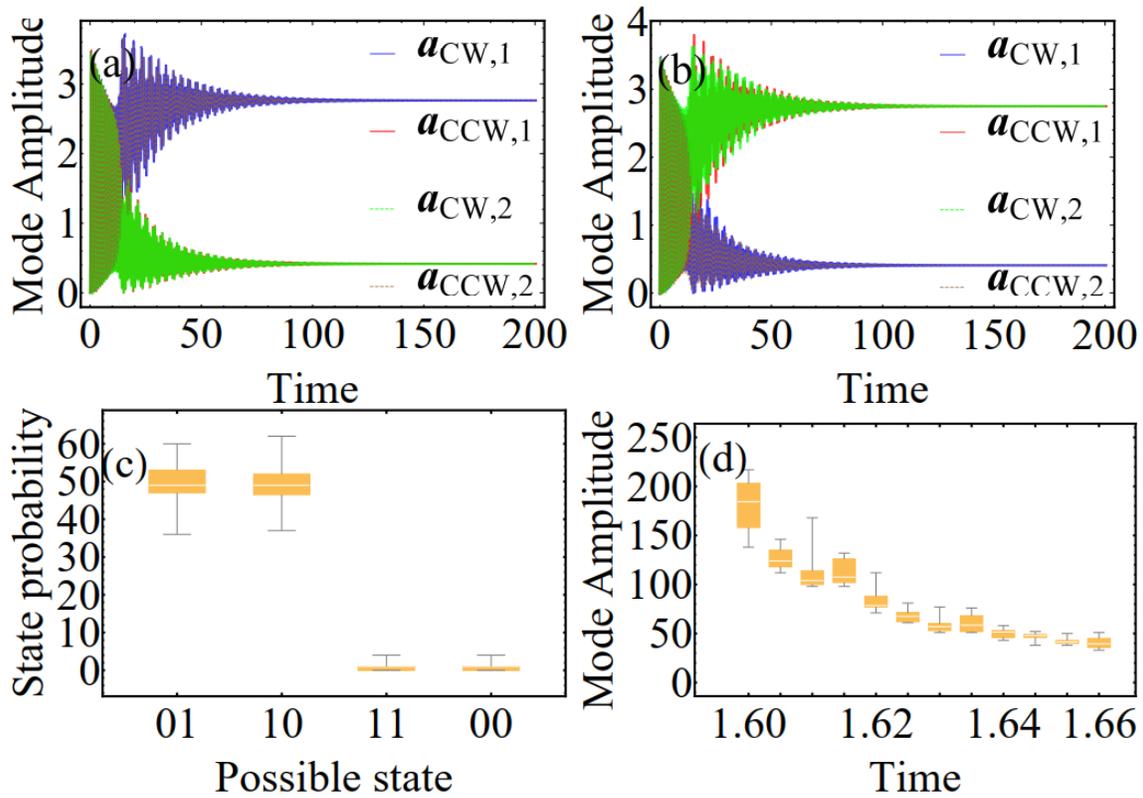

**Fig. 3. The dynamical behaviors of the coupled cavity.** The system is simulated with the pumping detuning as Δ=-1 , the dissipation as 0.1, the pumping power is 5. The scattering strength is g=1, the nonlinear coefficient is M=1. The cavities coupling strength between different cavities is $J_{ij}$=0.1. (a) and (b) represent the two main dynamic evolution modes in a dual-cavity coupled system. The total probability of the two modes is close to one . (c) Probability distribution of the four modes. (d) Statistical graph of the relaxation time under different pump powers.

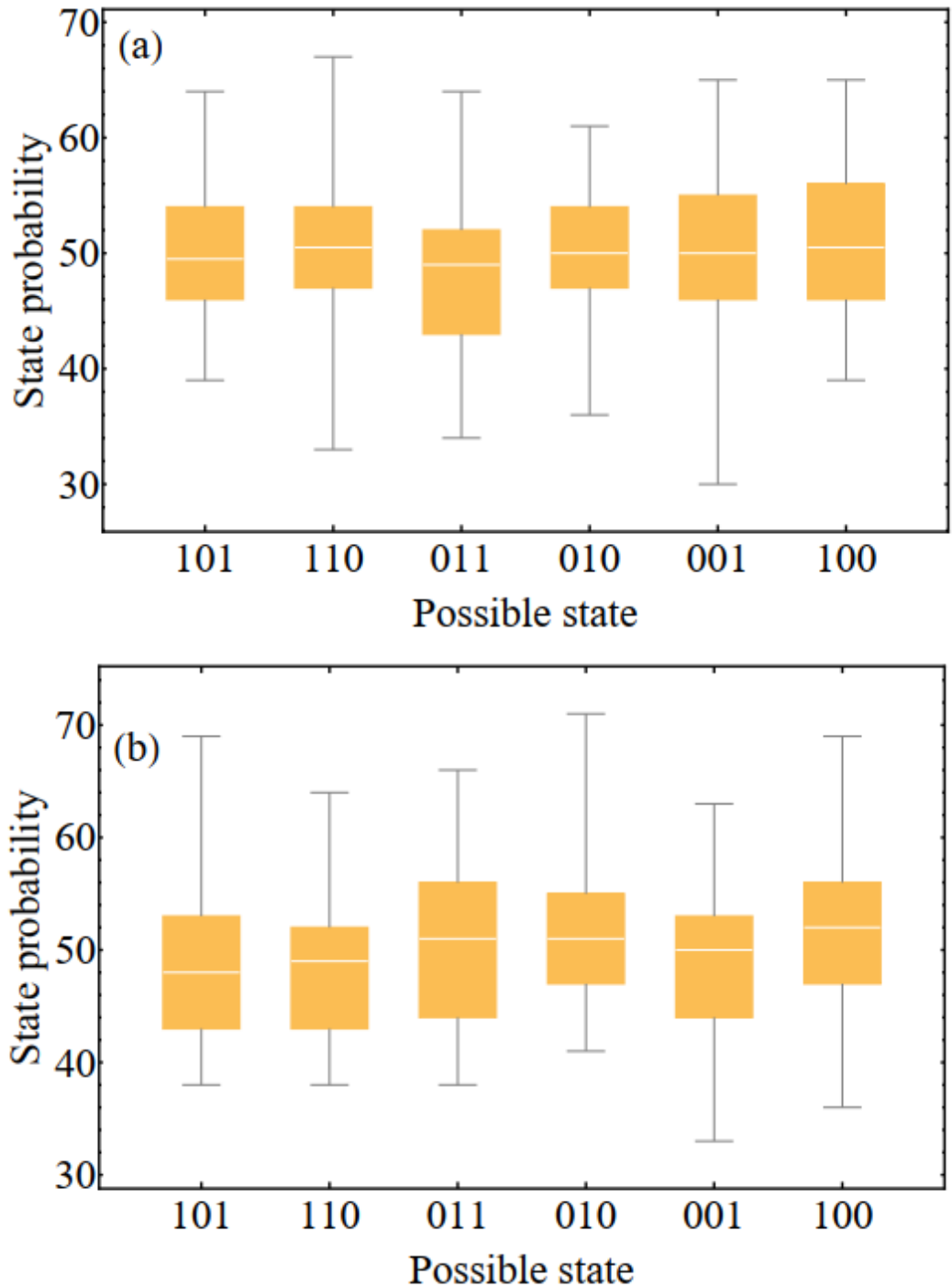

**Fig. 4. The state distribution of the three-body coupling under ferromagnetic coupling (a) and anti-ferromagnetic coupling (b).** The coupling strength between different cavities is 0.1. The drive detuning is -1, and the coupling strength between the CW and CCW modes in the cavity is 1, the nonlinear coefficient is 0.1. The drive strength is 5.

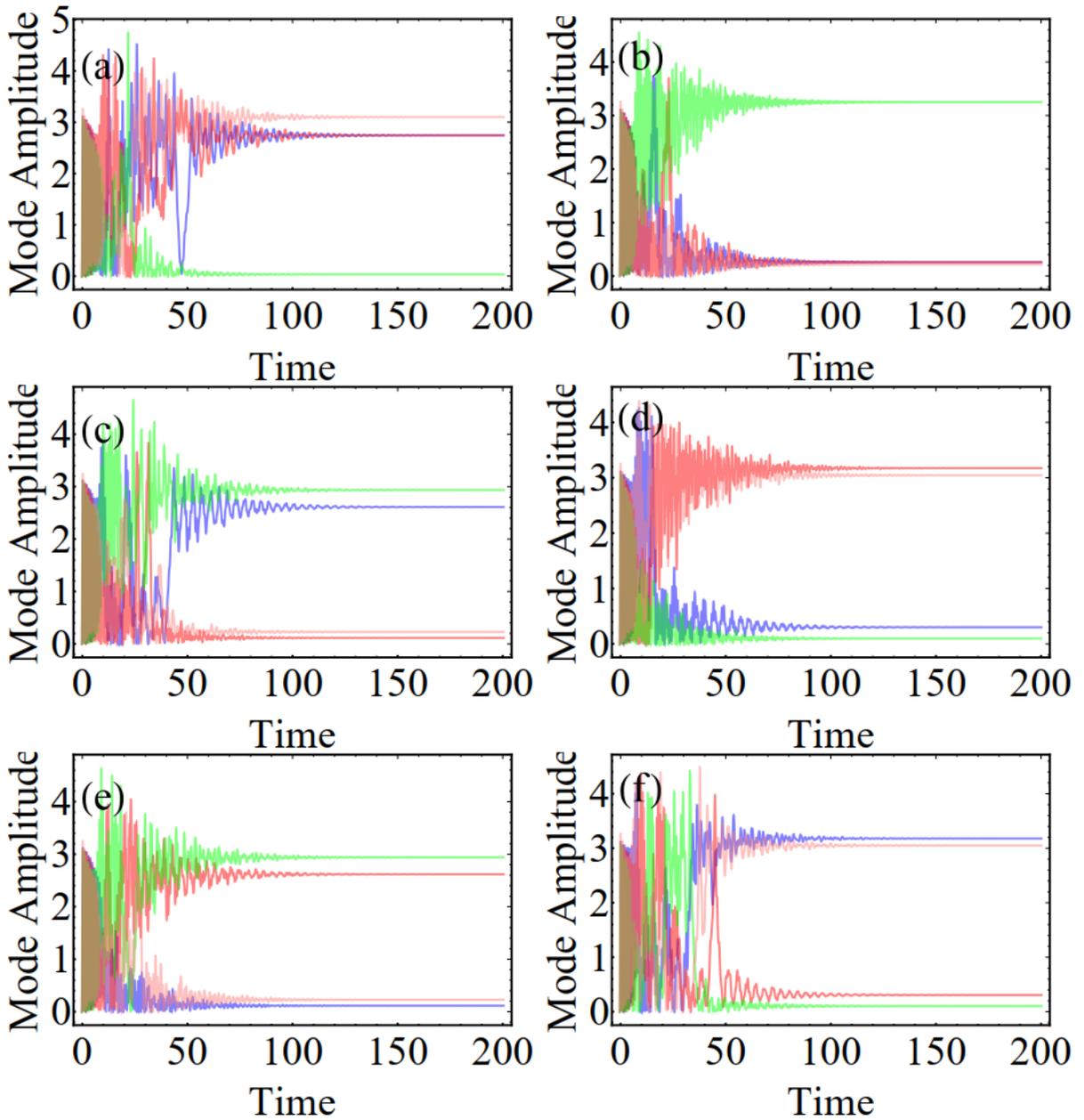

**Fig. 5. The simulation results of the three-cavity coupled in the annular form, while the fourth cavity coupled only with the third cavity.** The coupling strength between different cavities is 0.1. The drive detuning is -1, and the coupling strength between the CW mode and CCW mode in the same cavity is 1, the nonlinear coefficient is 0.1. The drive strength is 5. The dissipation rate of the cavity is 0.1. Here the blue line stands for the result in cavity-1, red line represents the result in cavity-2, the green line denotes the condition in cavity-3, and the pink line shows the results in cavity-4.

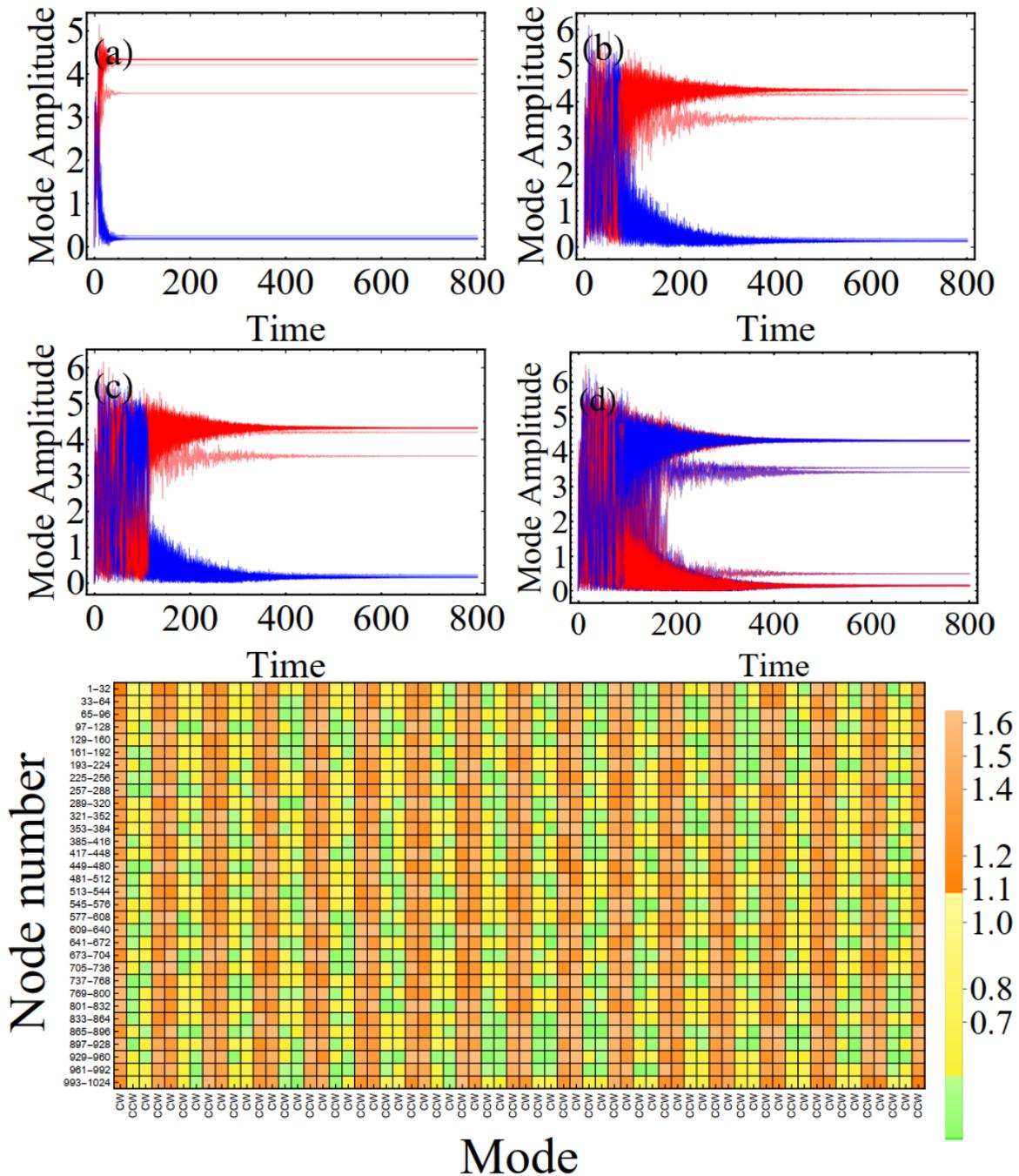

**Fig. 6. The results under the condition of large number of nodes.** The coupling strength between different cavities is 0.1. The drive detuning is -1, and the coupling strength between the CW and CCW modes in the same cavity is 1, the nonlinear coefficient is 0.1. The drive strength is 5. (a) The result for 16 qubits condition with damping rate 0.1. (b) The result for 32 qubits with damp 0.01. (c) The result for 64 qubits condition with damp 0.01. (d) The result for 256 qubits with damp 0.01. (e) The result for 1024 qubits. In order to show the results more clearly, we present the steady state results for each CW mode and CCW mode, and the colors correspond to the related amplitude of the modes.

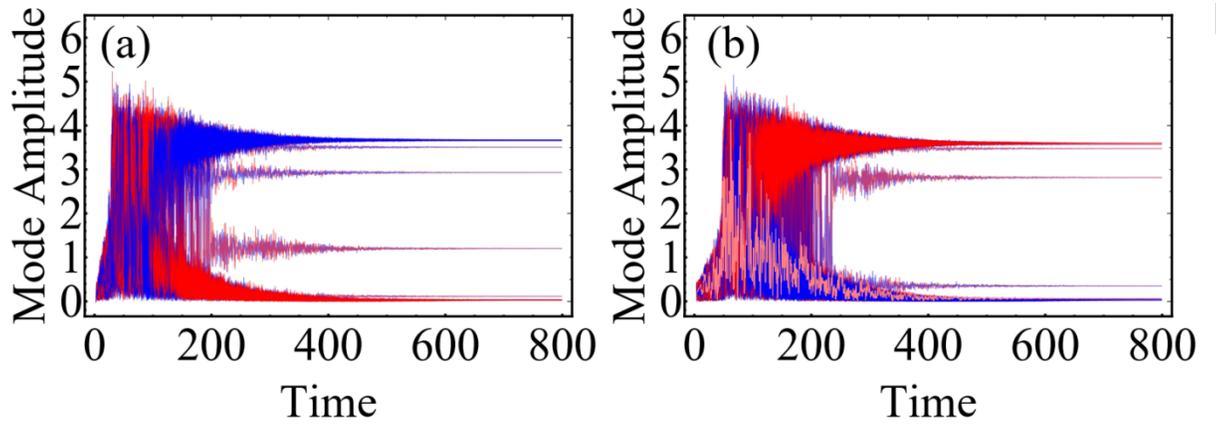

**Fig. 7.** The evolution of the field intensity of the 128 (a) and 256 (b) cavities system under calculation failure. The system is simulated with the pumping detuning Δ=-1, the dissipation rate 0.1, the pumping power is 5, the scattering strength is g=1, the nonlinear coefficient is M=1. The cavity coupling strength between different cavity is 0.1. The drive field detuning is -1, and the coupling strength between the CW and CCW modes in the same cavity is 1, the nonlinear coefficient is 1. The pump power is 5, the coupling strength between the CW and CCW modes in different cavities is 8. The damp of all cavities is 0.05.